\documentclass{KapProc} % Computer Modern font calls
 
\usepackage{graphicx}
\normallatexbib

\newcommand{\om}{\omega_m}

\newcommand{{ \vn }}{\vec{n}}

\newcommand{\beq}{\begin{equation}}
\newcommand{\eneq}{\end{equation}}

\newcommand{\met }{\frac{1}{2}}

\begin{document}

\articletitle{Compensation of the  spin of a quantum dot at Coulomb blockade }

\articlesubtitle{ Kondo correlation with the contacts 
leads to a macroscopic  separation of the charge from the spin in the dot}

\author{Domenico Giuliano}
\affil{ Department of Physics, Stanford University, Stanford, California 94305}
\email{asterion@partenope.stanford.edu}
\author{Benoit Jouault}
\affil{GES, UMR 5650, Universit\'e Montpellier $II$, 34095  Monpellier 
Cedex 5, France}
\email{jouault@ges.univ-montp2.fr}
\author{Arturo Tagliacozzo}
\affil{ Dipartimento di Scienze Fisiche, Universit\`a di Napoli 
        "Federico II ", Monte S.Angelo   via Cintia, I-80126 Napoli, Italy\\
 and       Istituto Nazionale di Fisica della Materia (INFM),
     Unit\'a di Napoli}
\email{arturo@na.infn.it}

\begin{keywords}
quantum dot, Kondo effect. PACS 71.10.Ay,72.15.Qm,73.23.-b,73.23.Hk,79.60.Jv,
73.61.-r
\end{keywords}

\begin{abstract}

We  discuss a new entangled state that has been observed in the conduction 
across a quantum dot. At Coulomb blockade, electrons from the contacts 
correlate strongly to those localized in  the dot, due to cotunneling 
processes. 
Because of the strong Coulomb repulsion on the dot, its
electron  number   is unchanged {\sl w.r.to} the dot in isolation,
but the total spin is fully or partly compensated. 
In a dot with  $N= \:\: even$    at the   singlet-triplet crossing, which  
occurs  in   large  magnetic field,   Kondo correlations lead to a  
total spin  $S=1/2$.
\end{abstract}

\section*{Introduction}

 A very special kind  of Macroscopic Quantum Coherence has  been recently
measured. This is the Kondo anomaly of the  conductance across 
 a   quantum dot (QD)  polarized at  Coulomb blockade (CB)
\cite{goldhaber,cronenwett,schmid}. 
Due to confinement, Coulomb correlations between electrons added to a
 QD  are strong. 
 An appropriate   choice of the gate voltage $V_g$
blocks  direct sequential tunneling, what is called  Coulomb blockade
regime\cite{kouwenhoven}.
If coupling between the  QD and the leads is not weak and  
the temperature is low enough,  an anomaly appears in the differential 
conductance  at zero voltage bias. Kondo correlations   are established 
 by cotunneling processes.  This leads to a macroscopic  ground 
state (GS) with  very peculiar properties, in which the dot and the 
contacts are entangled.

Kondo conduction is  a well known phenomenon occurring in non magnetic 
metals with  diluted  magnetic impurities\cite{hewson}. 
Typical examples are $Fe$  
in $Cu$  or $Mn$ in $Ag$:  
 a minimum in the resistivity of the metal occurs in lowering the 
temperature  below $T_K$, before   saturating   to  a residual  
value  corresponding to potential  (non magnetic ) scattering.
 Kondo showed that \cite{kondo}  there is a crossover to a correlated
state of the system 
in the neighborhood of  $T_K$, in which exchange interaction between the
 conduction electrons  and the local moment of the impurity 
leads to scattering 
events on the microscopic scale, in which the electronic spin is 
flipped.  This gives rise to a term in the impurity contribution 
to the resistivity that increases with decreasing temperature.
Spin flip can only occur if  the spin on the local magnetic 
moment of the  impurity  atom changes  accordingly, to achieve 
compensation.
 This shows up in   the disappearance of a Curie-Weiss 
spin susceptibility  which becomes constant at low  temperature.

To some extent  the conductance across a QD 
in the CB  regime  can be regarded as  the  
mesoscopic realization of the same physics\cite{theo}.
This is not so surprising, as  QD have always been referred to 
as artificial atoms and the contacts provide the delocalized conduction 
electrons  of the host  metal.

 Of course the energy scales are quite different. 
Energy level separation in a QD is of the order of $meV$ while it is of 
two or three orders of magnitude larger for the $d,f$ electronic levels in   
the impurity atom. This implies that the temperature scale 
(the Kondo 
temperature  $T_K$ ) is correspondingly
reduced from  tens of $K$ to   $100 mK$  and below.
Also, because the anomalous contribution to the conductance stems from
cotunneling processes, the internal structure of the QD is of big 
relevance. Coulomb interaction is  dominant  in QD, so that the  levels
involved are many particle levels which strongly depend on the 
number of electrons  added to the dot $N$ by means of a gate voltage $V_g$.
 
In this sense one should be cautious  when  extending 
 the single impurity
Anderson model  to the dot case\cite{anderson}. 
The model  describes  localized single 
electron levels 
with an   onsite Coulomb repulsion $U$,  in hybridization  with 
delocalized conduction electrons. It is important that the order of magnitude 
of $U$ is the same for  both the magnetic atom and the QD, so that, in the 
dot case, we are always in the large $U$ limit, what  makes the 
 experimental observation   even more striking. In particular, the Kondo 
peak is  observed  in a QD in the CB region where conductance is otherwise 
exponentially  small. This is because 
QDs   can be tuned within a very wide   range  of parameter values 
and the experiment on Kondo  conduction is  equivalent to the measurement  
of one single  impurity  in the metal host. 

As we will show, the investigation of  the  analogies between the two 
systems is
  very fruitful,  because new features arise in QD, that  are not found 
when localized moments in diluted alloys  are studied.

One example is  Kondo conduction in a magnetic field\cite{silvano}. 
Because orbital effects are very important in a QD when a magnetic field is 
applied, the energy scale for the Zeeman spin splitting is not the same 
as the energy scale for level separation in a QD in magnetic field. 
This implies that, in diluted alloys, a magnetic field $B$ is always disruptive for 
Kondo correlations
because it lifts the degeneracy of  the levels, which is required for the 
spin flip scattering events. On the dot side,  it may even produce 
crossings of  the  levels of the dot,  which  could favour Kondo 
conductance. We will discuss an example of  this in Sect.  2.    

The spectroscopy of a QD is usually performed in terms
of transport measurements\cite{tarucha}. Two electrodes are attached to the QD and the
low-temperature zero-bias conductance
is monitored. When the gate voltage $V_g$ changes, the conductance 
undergoes   a  series of peaks at zero source-drain  voltage $V_{sd}$ 
each time the increase in  $V_g$ matches the  chemical potential for adding
an extra electron to the dot (Coulomb oscillations).
In between two peaks, the number of particles on the QD is fixed
 (CB regime)( see Fig. 1  for the setup $(a)$ and for a schematic 
grey-scale drawing of the conductance $vs$ $V_g$ and $V_{sd}$ $ (b)$) . 
Contributions to the current at CB 
are fourth order in the transmission amplitude  and can be very small
(white regions in Fig.1 b)).
In these conditions the total spin of the dot $S$ is the only left 
dynamical variable, as in a magnetic impurity.   

Provided this spin is not a singlet, it can become strongly coupled to 
the spin density of the contacts electrons if $T < T_K $
  and  the leads are  not weakly linked. 
The correlated state gives rise to non perturbative differential 
conduction at zero 
voltage bias, fairly independent of the value of $V_g$ within the 
CB plateaux. 
\begin{figure}[ht]
\centering
\includegraphics*[width=4in]{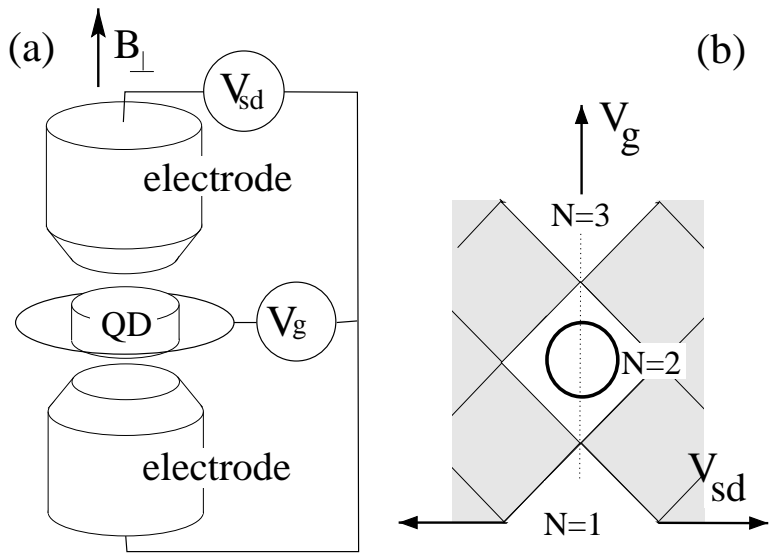}
\vskip.2in
\caption{$a)$ Geometry of a vertical QD. $b)$ grey scale 
picture of the conductance versus gate voltage $V_g$ and source-drain 
voltage $V_{sd}$. In white areas the conductance is vanishingly small.}
\end{figure}

The first observations were in dots at CB with an odd number of electrons
\cite{goldhaber,cronenwett}. 
The GS is a doublet and coupling to the contacts in the CB region 
generates the Abrikosov-Suhl resonance at the Fermi energy $\mu \equiv 
\epsilon _F$ of the contacts
 which gives a peak in the differential conductance at zero bias
due to cotunneling across the QD.

 In Sect. 1  we review   the features  of 
  the single impurity 
Anderson model which  are at the basis of the physics involved in the 
Kondo correlated state for  QDs  with an odd number of electrons. 
The level of the localized impurity is doubly degenerate 
and located deeply below  $\mu$  but double occupancy of this level would 
cost an energy much larger than $\mu $.  We show  that,
   in the limit 
of strong  onsite  repulsion $U$, in spite of the coupling to the leads,
 occupation of the   impurity level is frozen. 
In addition to  the disappearance of   the charge  degree of freedom, 
a singlet state is  generated  on the impurity, with the help 
of   the conduction electrons. 
  In the symmetrical case,   the fixed point  GS 
 has $(N=1, S=0$).  Spin-charge separation occurs.  
 
Dots   at CB with an even number of electrons are not expected  to give 
Kondo  conduction because the GS is supposed to be  a singlet already.
A notable exception is when the GS has higher total spin. 
 In dots with $N=4$ and $N=6$ 
Hund's rule states that the GS  is a triplet. This is   turned into a singlet 
by 
applying a small $B$ ( denoted ``TS crossing'' hereafter). 

 In Sect. 2 we  consider the case 
of a realistic isolated QD with few electrons using 
exact diagonalization methods 
\cite{jouault}.  A
 magnetic field $B_\perp$, orthogonal to the dot is applied. 
This produces large  orbital changes in the electronic states and, 
eventually,  crossing of levels. 
The Kondo effect expected for $S=1$ 
is strongly  enhanced due to this crossing and the 
anomalous conductance at zero voltage has been recently observed
\cite{silvano}. The very peculiar physics at the TS crossing is 
under  consideration at present\cite{eto,pg}.

\begin{figure}[ht]
\centering
\includegraphics*[width=4in]{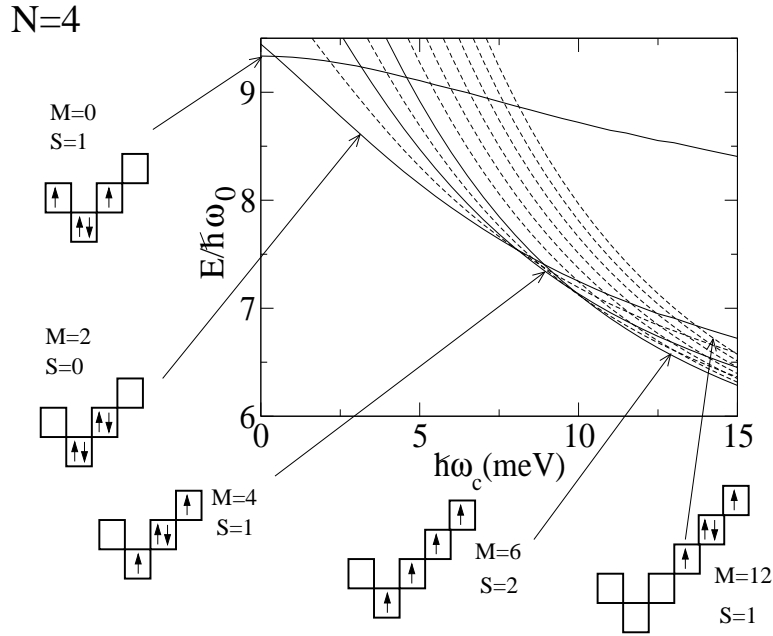}
\vskip.2in
\caption{Energy levels of an isolated QD with $N=4$ electrons $vs.$ 
a magnetic field orthogonal to the dot. The total orbital angular 
momentum and the total spin of the different GSs are shown, together 
with the configuration  of the Slater determinant which has the highest
projection onto the GS.}
\end{figure}

But there is another possibility of Kondo conductance 
which has been studied in ref\cite{noi}.
Increasing  $B$ quite substantially, the dot undergoes a new  transition
 from the singlet to triplet state because of orbital effects (
denoted  ST crossing  hereafter),  in which the  Zeeman spin splitting cannot
be ignored. This changes the properties of the GS in a peculiar way,
leading to an even charge on the dot with total spin $\met$.    

\section{From the symmetric Anderson model to the impurity spin dynamics}

We refer to a  QD connected 
to the leads in the CB regime, in   the limit 
of the Coulomb interaction being very large. 
As discussed in the introduction,
  the dot charge degree of 
freedom is frozen  out, what
leaves  the dot total spin $S$ as the only dynamical variable. The  
charge-spin separation  emerges. This  phenomenon is at the basis of  
 the analogy between this system and that of an impurity 
local moment in a metal alloy. 
Here we review the main features of the single impurity spin $\met$ 
symmetric Anderson model\cite{anderson}. The path integral formalism by 
 Anderson,Yuval and Hamann \cite{yuval,hamann} is particularly suited 
 to focus on the spin-charge separation in this  model. 

 The Hilbert space of the  spin $\met $ impurity is conveniently described 
by two  fermionic operators, $d_{i}^\dagger$,  acting on the 
vacuum  $ | 0 \rangle$.   The label $i =1,2$ is here a spin label, 
although it  could be a generic label for the two degenerate 
single particle levels located at energy $\epsilon _d$.

The impurity Hamiltonian is:

\beq
H_{ d} = \epsilon _d \sum_i n_i + U n_1 n_2
\label{heff}
\eneq
where $n_i =d^\dagger_i d_i $ and $U$ is the strength of the onsite 
repulsion.

We linearize the conduction electron  band  $\epsilon_{ k }$  
close to $\epsilon _F $. This 
 allows  to write down their Hamiltonian as that of a one-dimensional 
Fermi gas. To keep contact with the dot picture, we  divide  the space
into a left $(L) $ and a right $(R)$ region:
\beq
H_{ L , R } = \sum_{ k \sigma} \epsilon_{ k}  b_{ (L, R) k \sigma}^\dagger
b_{ ( L , R ) k \sigma}
\label{leads}
\eneq
 If  $\epsilon _d = -U/2 $ $w.r.to $ the chemical potential
 of the conduction electrons $\mu$
(taken as the zero of the single particle energies),
  the Anderson model which arises is 
symmetric. In fact, the energies  of the empty impurity state,$^0E$,
 and that of 
the doubly occupied impurity state, $^2E = 2 \epsilon _d +U $,
 are both  zero, while 
the singly occupied impurity level has energy $^1E = -U/2$.     

Finally, the hybridization  Hamiltonian between the conduction electrons 
 and the impurity is:

\beq
H_t = \frac{1}{ \sqrt{2}} \sum_k \{ \Gamma^* [ ( b_{ L k \uparrow }^\dagger +
b_{ R k \uparrow}^\dagger ) d_1 + ( b_{ L k \downarrow }^\dagger +
b_{ R k \downarrow}^\dagger ) d_2 ] + { \rm h.c.} \}
\label{tunnel}
\eneq

where all the tunneling amplitudes have been taken equal for simplicity.

We now formulate the quantum dynamics of the system described by eqs.(
\ref{heff},\ref{leads},\ref{tunnel}) using the  imaginary time Feynman Path
Integral formalism. Following \cite{hamann}, we  
integrate out the conduction electron fields and we obtain the partition function in terms of the Grassman fields $d, d^\dagger$ only
($\hbar = 1$  in the following):

\beq
{\cal{Z}}(\mu)  \propto
\int  \Pi_i \left ( D d_i D d^\dagger_i \right )  e^{  -  {\cal {A}}_d } \:
 e^{-\beta |\Gamma |^2 \sum_{\om, i } d^\dagger _i K(i\om )d_i}
\eneq
where  $i \omega _m $ are Fermionic Matsubara frequencies,
\beq
{\cal {A}}_d = \int_0^{ \beta } d \tau
\left \{ \sum_{i} \left [ d_{i}^{\dagger}
\frac{\partial}{\partial \tau } d_{i}+
 \epsilon_d \: d^{\dagger}_{i}
d_{i} \right ] + U d^{\dagger}_1 d_1 d^{\dagger}_2 d_2
\right \}
\eneq
and 
$ K(i\om ) $ is:
\beq
 K(i\om ) = \frac{L}{2\pi}\int _{-D}^{D}\frac{dk}{  i\om +vk} =
\frac{ L}{2\pi v_F }
\ln  \left (-\frac{v_FD+i\om}{v_FD-i\om}\right ) \:  .
\eneq 
Here $v_F$ is the Fermi velocity and $D$ is the band cutoff, symmetrical 
$w.r.to \: \epsilon _F$. 

The average impurity occupation number is calculated from the retarded 
part of the Green function $G^{R}(\omega )$:
\begin{eqnarray}
<n_{i} > = \int _{-\infty}^{\infty}  \frac{d\omega}{2\pi} i 
G^{R}(\omega )\: . 
\nonumber
\end{eqnarray}
 $G^{R}(\omega )$ is obtained from $G(i\om )$  in the 
 limit to real frequencies: $ i\om \to \omega + i 0^+ $.  
In the case $U =0$ we have:
\begin{eqnarray}
G _{(0)}^{-1} (i \omega _m ) = i \omega _m  -\epsilon _d + 
|\Gamma |^2  K (i\om) \to  \omega -\epsilon_d -\frac{\Delta }{\pi} 
\ln  \left |\frac{v_FD+\omega }{v_FD-\omega}\right | + i \Delta 
\nonumber
\end{eqnarray}
where  $\Delta = \pi N(0) |\Gamma |^2 $ and $N(0)$ is the density of states 
at the  Fermi energy per spin $L/2\pi \hbar v_F$.

 Assuming that 
$\tilde{\epsilon} _d $ solves the equation $ \Re e \{ (G^0 )^{-1} \} =0$,
we get:
\begin{eqnarray}
<n_{i} > = \int _{-\infty}^{\infty}  \frac{d\omega}{2\pi} i 
G_0^{R}(\omega ) 
\approx  i\int _{-\infty}^{\infty}  \frac{d\omega}{2\pi} \:
\frac{1}{\omega -\tilde{\epsilon} _d + i \Delta }
\nonumber 
\end{eqnarray}
The imaginary part is odd and vanishes upon integration. The real part gives:
\beq
<n_{i} > = \int _{-\infty}^{\infty}  \frac{d\omega}{\pi}  
 \frac{\Delta}{(\omega -\tilde{\epsilon} _d)^2 +  \Delta ^2 } =
\frac{1}{2} - \frac{1}{\pi} \arctan \frac{\tilde{\epsilon} _d}{\Delta }
\label{occupa1}
\eneq
 Eq.(\ref{occupa1})
describes   the Friedel screening around the impurity due to a resonance at 
 $\tilde{\epsilon _d } \approx \epsilon _d$  of width $\Delta$.  

We now discuss the role of the onsite Coulomb interaction.
In the large-$U$ limit we have:
\begin{eqnarray}
\exp  \int_0^\beta d\tau \left \{-\epsilon _d(n_1+n_2)-Un_1n_2\right \}
 \nonumber\\
= e^{\frac{\beta}{U}\epsilon _d^2}\; \delta (n_1+n_2 
+ 2\epsilon _d/U)
\: \cdot e^{\frac{U}{4}\int_0^\beta \;d\tau (n_1-n_2)^2} \:\:,
\label{hub}
\end{eqnarray}
where the delta function implements the constraint of single site 
occupancy  in the symmetric case,  $\epsilon _d= - U/2 $.

The quartic interaction is decoupled by means of a Hubbard-Stratonovitch
boson field $X ( \tau )$, according to the identity:

\[
e^{ \frac{U}{4} \int_0^\beta d \tau ( n_1 - n_2 )^2 } =
\int D X e^{ - \frac{1}{4 U} \int_0^\beta d \tau ( X^2 ( \tau )
+ 2U ( n_1 - n_2 ) X ( \tau )) }.
\]
Having introduced $X ( \tau )$, ${\cal Z} ( \mu )$ takes the form:

\begin{eqnarray}
{\cal{Z}}(\mu) \propto
\int D X e^{ - \frac{1}{4U} \int_0^\beta d \tau  X^2 ( \tau )}\nonumber\\
\times
 \int   \Pi_i \left ( Dd_i Dd_i^\dagger 
 e^{ - \int_0^\beta d \tau d \tau ' d_i^\dagger ( \tau )
G^{-1}_{(U)}(\tau -\tau ')d_i (\tau ')} \right )  
\nonumber \\ \times \delta (n_1+n_2 -1) 
 \cdot \met \sum_{j=1,2} e^{  (-1)^{j}\int _0^\beta d \tau 
 [ n_j - \met ] X ( \tau )} \:\: .
\label{part2}
\end{eqnarray}
Note that now the term $ \epsilon _d \sum _i n_i $ was included in
 eq.(\ref{hub}),
so that   $G^{-1}_{(U)}(i\om) = i\om +|\Gamma |^2 K(i\om)$ in this case.
At odds with  the case $U=0$ here we have $\tilde{\epsilon _d } \approx 0$. 
The resonance is at the Fermi level, in spite of the fact that the 
original localized level is at $\epsilon_d $. This makes 
eq.(\ref{occupa1})
consistent  with the single site occupancy constraint.   
The partition function in eq.(\ref{part2}) describes an effective spin-1/2
coupled to the fluctuating magnetic  field $X ( \tau )$. Its 
  dynamics is constrained by the requirement that the impurity is 
singly occupied. This should be implemented at any imaginary time, 
what is quite hard \cite{michela}. In the following it will be accounted 
for in  the 
average. Indeed,  the average   over the  $(0,\beta )$ interval,
of the field configurations we  consider,
 satisfies it.    

Therefore, putting aside the delta function, the  integration to be 
performed  
over the Grassman fields  $d_i, d_i^\dagger$  is:

\begin{equation}
\int D d_i D d_i^\dagger e^{ \int_0^\beta d \tau d
\tau^{'} d_i^\dagger ( \tau )  G_{(U)}^{-1} ( \tau - \tau^{'} )
 d_i ( \tau^{'} )  
 - \int_0^\beta d \tau X( \tau  ) [ n_  {i }( \tau  )
-  \frac{ 1}{ 2}  ] }
\label{efx1}
\end{equation}

By including  a coupling constant $g$
in front of $X$  we obtain the result:
\begin{eqnarray}
e^{ - \int_0^1  dg \int _0^{\beta} X (\tau )  
[ G ( gX , \tau, \tau^+ )-  \frac{ 1}{ 2} ] } d\tau
\nonumber
\end{eqnarray}

Because 
$ G_{(U)} ( \tau  ) \approx -\frac{ 1}{ 2 \pi \Delta} \:\:{\cal P}
(1 /\tau ) $, 
the equation for  $G [ \xi , \tau
, \tau^{'} ] $:
\begin{equation}
G (\xi, \tau , \tau^{'} ) = G_{(U)} ( \tau - \tau^{'} ) +
\int_0^\beta d \tau^{"} G_{(U)} ( \tau - \tau^{"} ) \xi ( \tau^{"} )
G (\xi, \tau^{"} , \tau^{'} )
\label{mus}
\end{equation}
 is  the Mushelishivili equation \cite{hamann}. 
 $G$ may be split into two contributions, one
that  is 
singular as $\tau^{'} \rightarrow \tau$, and a  second one that is
regular. The singular contribution is given by:

\begin{equation}
- \frac{ 1}{ \Delta } \frac{1}{ 1 + \xi^2 ( \tau )} \left[ \frac{1}{ \pi}
{\cal P} \frac{1}{ \tau - \tau^{'} } + \xi \Delta \delta ( \tau - \tau^{'}
) \right]
\label{sing}
\end{equation}
where $\xi (\tau ) = X(\tau ) / \Delta $.

Eq.(\ref{sing}) generates an effective potential for the field $\xi$,
$ V [ \xi ]$, given by:

\beq
V[ \xi ] = 
\int_0^\beta d \tau \Delta  \left[ \frac{\Delta }{ 4U } \xi^2 - 
 \frac{1}{\pi} \left ( \xi 
\arctan ( \xi ) - \frac{1}{2} \ln (1+\xi ^2 ) \right )\right]
\label{pot2}
\end{equation}

If the condition $ \Delta / 4 U  < 1 $
is satisfied, $V$ has non-trivial minima  at 
$ \xi_0 = \frac{ 2U }{\pi \Delta } \arctan (\xi_0) 
 \approx \pm   U /  \Delta$  ( for large $\xi _0$ ).

The saddle point solution  $\xi_{sp} ( \tau )$ is  either static  or, if
it is $\tau -$dependent, it is periodic. It  represents  a series of jumps
between the two minima. All these solutions satisfy the constraint on the
average within the imaginary time interval $(0,\beta )$, because:
$ \langle n_i -\met \rangle \approx - (-1)^{i}  
\overline{X_{sp}(\tau )}/2U =0$.

In the next section we shall map the system onto a 1-dimensional Coulomb Gas
(1-d CG) of jumps between the two minima (instantons) and shall derive the
low-temperature Kondo physics as condensation of instantons by means of a
Renormalization Group (RG) analysis.

\subsection{The equivalent 1-dimensional Coulomb Gas.}

The mapping between the low-temperature Anderson model and a 1-d CG of
instantons has been extensively discussed in the literature \cite{yuval,nie}. 
Hence,
here we will just skip all the steps leading to the final form of the effective
action. The full partition function may be approximated with the sum over the
trajectories given by hopping paths and will be given by:

\begin{equation}
Z = \sum_{ N = 0}^\infty \frac{1}{ 2N ! } \int_0^\beta d\left (\frac{ 
 \tau_1 }{
\tau_0 }\right )  \ldots \int_0^\beta d \left (\frac{ \tau_{ 2N}}{ \tau_0 }
\right ) [ e^{ 
\frac{1}{2}\sum_{i
\neq j = 1}^{2N} ( -1 )^{i+j} \alpha^2 \ln \left| \frac{ \tau_i - \tau_j}{
\tau_0} \right| } Y^{ 2N} ]
\label{part1}
\end{equation}

The first relevant parameter is the ``fugacity" $Y$, corresponding to
the probability for an istanton to take place within  a certain state of the
system. It is given by:
\[
Y = \tau_0 e^{ - \bar{{ \cal A}}}
\]
where $\bar{ \cal A} \sim \tau_0 U$ is the action of the single blip.

The second parameter is provided by the bare ``interaction strength" between
instantons:
\beq
\alpha _b = \sqrt{2} (2 \arctan(\xi _0) )= \sqrt{2}
(1- 2\Delta/\pi U)
\label{alpb}
\eneq
The integral over the ``centers of the Instantons" has to be understood
such that $\tau_i$ and $\tau_j$ never become closer than $\tau_0$.
 Here the instantons interact via a
logarithmic potential $ \ln | (\tau _l - \tau _{l'} )/\tau_0 |$, 
so that  an ultraviolet cutoff, $\tau_0$ is needed. The requirement 
that  the physics is  independent of  $\tau_0$ provides the RG 
equation for the parameters. By  rescaling  $\tau_0$ to $\tau_0 + 
d \tau_0$,  one obtains the
scaling of the fugacity and the renormalization of the coupling
constant induced by processes of fusion of charges, which
 lead to the renormalization group equations \cite{nie}:
\beq
 \frac{ d Y}{ d \ln \tau_0 } = ( 1 - \frac{\alpha ^2}{2}) Y
\; ; \frac{ d \alpha ^2}{ d \ln \tau _0 } =  - 2 Y^2 \alpha ^2
\label{rg1}
\eneq
The scaling eq.s (\ref{rg1})  are discussed in the next subsection. 
Here it is enough to mention that, according to eq.(\ref{alpb})  
 the regime relevant to our analysis has  $\alpha^2/2 < 1$. Then,
the flow is towards $ Y \to \infty $ and $\alpha ^2 \to 0 $. The corresponding
phase is characterized by a ``proliferation" of instantons, namely, by a
continuous cotunneling  between the impurity  and the leads.
Because of the antiferromagnetic coupling (AF), 
this produces a continuous flipping 
of the  impurity spin  $S_{eff}$  (whose $z-$ component 
is $\langle (n_2  -n_1)(\tau)  \rangle $). The latter  is screened out,
what  implies 
charge-spin  separation:  the charge on the impurity is 1,
 while the spin is zero.  We  estimate now the Kondo temperature,
$T_K$, by using the RG equations (\ref{rg1}).

\subsection{The Kondo temperature.}

The Kondo temperature   $T_K$ is
defined as the temperature at which   instanton 
condensation  sets in. In order to estimate it, we use the system of
equations (\ref{rg1}), starting the flow from   $ \Delta/ U
  \ll 1$. 
If we   define $\chi = 1 - \alpha^2 /2$, 
eq.s (\ref{rg1}) can be approximated as:
\beq
\frac{ d Y^2 }{d \ln \tau_0 } = 2 \chi Y^2 \;\;\; ; \; 
\frac{ d \chi^2 }{ d \ln \tau _0 } = 2 \chi Y^2 
\label{rg2}
\eneq
The system (\ref{rg2}) has a constant of motion, given by $
Y^2 - \chi^2 /2 = Y_0^2 - \chi_0^2 /2 = cnst $.
A particular case    is when  $Y_0 = | \chi_0 |/\sqrt{2}$.
 Its RG trajectory 
is  a straight line in the $Y$-$\chi$ plane, which separates two phases:

 $i)$  The phase with $| \chi_0 |/\sqrt{2} < Y_0$ and  $\chi_0 < 0$, that is 
attracted by the line ($Y_\infty = 0$, $\chi_\infty = cnst$  $<$ 0 );

 $ii)$  The phase  with  $| \chi_0 | /\sqrt{2} > Y_0$, $ 
or $  $\chi_0 > 0 $,   that is attracted
by the point ($Y_\infty = \infty$, $\chi_\infty = \infty$).

It should be kept in mind, however,  that eq.s(\ref{rg1},\ref{rg2}) 
are only valid 
in the first stages of scaling, so that these flow limits   do 
not strictly hold, as demostrated by the exact solution \cite{tsvelick}.
In any case, the line   $Y_0 =
 \chi_0 /\sqrt{2}$ signals the crossover between the
two regimes.  The equation for  $\chi$ on this line becomes:
$ d \chi / d \ln \tau _0  = \chi^2 $.  According to this equation 
$\tau _0 \exp ( 1/\chi ) $ is a scale invariant.  Because 
a decrease of  temperature described by $d\beta$ corresponds to a dilation of 
$\tau _0$ according to $  d \tau_0 / \tau_0  =  d \beta / \beta  $,
  the scale invariant temperature is  
$ T_K =  \tau _0^{-1} e^{ - 1/\chi _0} = 
 \tau _0^{-1} e^{ - (\pi U)/( 4 \Delta )} $.

The prefactor $ \tau _0^{-1}  \approx ( U\Delta )^{\met}$
 can only be obtained by including two loop  corrections. 

In the language of the Coulomb gas of instantons, $T_K$ can be interpreted as
the temperature where the screening of the effective interaction 
between charges starts. Below this temperature 
$Y$ flows to infinity and  $\alpha^2 $ flows to zero.
 This  corresponds
to a fully coherent state, the ``instanton condensate",
 in which the impurity spin   is fully compensated.

\section{ Kondo effect in magnetic field}

In Sect. 1 we reviewed the Anderson model for the electronic state of 
a localized  impurity  that hybridizes with a continuum of conduction 
electrons. This model   can be adopted  as a paradigm to describe  Kondo 
conduction in a QD,  provided some extra features of the QD (which is 
certainly  much a more complex object than a magnetic impurity atom)  
are accounted  for.   Clarification of 
these points also offers  the way through to  some peculiarities of 
the Kondo state in QDs that are absent in  the conduction of 
 diluted alloys.  Indeed, astonishing properties have been experimentally 
tested of  the  Kondo effect in a QD in magnetic field 
\cite{silvano} and we believe that 
others will  be soon revealed. 

Here we quote  the most relevant ones. 

 Because  the dot has an internal 
structure  its states are many body states which  cannot be obtained by 
application of  fermionic single particle operators $d_i$ as we did in 
Sect. 1. Only under very special circumstances this is possible, in an
approximated way\cite{noi}.
 The role of the excited states should also be considered.
One more aspect to be taken care of is the fact that, being the dot 
not point-like, the symmetry of the coupling to the leads 
can be  important. In this respect a vertical geometry with azimutal 
symmetry, as the one sketched in Fig. 1 (a), offers some  control. 
In this case the angular momentum is conserved in the tunneling from
the contacts to the dot and the theory is effectively one-dimensional,
as in Sect.1. 

The scenario of Sect.1 can be applied to dots at CB 
 with an odd number of electrons,
whose  GS  is a doublet ($S=\met $). 

In the case of $N$ being even, the GS is expected to be a singlet, what 
would rule out the possibility of Kondo coupling. Indeed first observations
 in the absence of magnetic field reported a ``parity `` effect, by which 
Kondo behaviour was alternating with increasing $N$\cite{cronenwett}.

However, orbital effects are very strong in a QD,  when a magnetic field 
orthogonal to the dot plane, $B_\perp$, is applied. This could produce 
higher spin states and crossing  of levels, which give rise to Kondo 
conduction also  when $N$ is even. 

Figure 2 shows the many body energy levels of an isolated dot confined by a 
parabolic potential in a magnetic field $B_\perp$ ($\omega _c = eB/mc$) for 
$N=4$. Quantum numbers are the total angular momentum along the 
$z-$direction $M$, the total spin $S$ and the spin component $S_z$.
The configuration sketched aside represent the filling of single 
particle orbitals ( which are $2d$ harmonic oscillator orbitals: 
 $n$ is an integer,  $m= -n,..n$ is the orbital angular momentum, increasing 
by steps  of two) in the Slater determinant which 
has largest weight  in each state ($M= \sum _{i=1,N} m_i $, $S_z = 
 \sum _{i=1,N} \sigma _i $ where $m_i $ and $\sigma _i $ are the  angular 
momentum and spin component  of each electron along $z$).

At zero magnetic field  Hund's rule applies and the GS  has $M=0$ and $S=1$. 
Even a small $B_\perp$ generates large orbital changes in the electronic 
state.  Because of $B$, Hund's rule breaks 
down and the GS  becomes  a singlet, so that triplet and singlet levels cross 
( the TS crossing).  The Kondo effect expected for $S=1$ 
is strongly  enhanced due to this crossing and the 
anomalous conductance at zero voltage has been recently observed
\cite{silvano}. The very peculiar physics at the TS crossing has been 
also theoretically studied \cite{eto,pg}.

By increasing $B$ further, other  level crossings are met
(few  of which appear in Fig.2). They correspond to an increase of the 
orbital angular momentum with magnetic field, and possibly of the 
total spin.   The first of these crossings
is the one between  ($M=2,S=0$)  and ($M=4,S=1$) which occurs at a 
value of the magnetic field which is quite substantial ( the ST point).

Occurrence of  the ST point is quite generic in dots with $N=\: even$.
     In fact, with increasing of $B$,  $M$ increases to take advantage 
of the Zeeman orbital term and to reduce  the Coulomb 
interaction whose strength increases also. 
Meanwhile, the  total spin increases  up to the largest possible value, 
thus producing a gain in exchange  energy.
Zeeman spin splitting, being 
a small correction,  is not included.

The prototype of such a crossing is the Singlet-Triplet crossing for $N=2$, 
 we focus on  in the following. 
The field value is  $B_*  \sim 4 T $, but it can be modulated over a 
wide range. 
The GS for $N=2$ in the absence of $B_\perp$ 
is a non degenerate spin singlet, $~^2 S^0_0$. At 
$B_\perp = B_*$, if Zeeman spin splitting is sizeable, first crossing occurs 
between  $~^2 S_0^0$ and  the spin
triplet with total angular momentum $M=1$,  $ ^2 T^1_1$ ( $ ^2T_{S_z}^M 
: S=1, M=1 $).  It is important that,  at $B=B_*$, the  total spin of 
the dot  is the only  dynamical variable. This can be inferred  from Fig. 3, 
where the charge and 
spin density are plotted for the $N=2$  dot at $B=B_*$. While the charge 
density $\rho (r) $ in the dot is unaffected when $B$ moves across $B_*$, 
the spin density jumps 
dramatically from zero when is  $B<B_*$  to 
$\sigma ^z (r) = \frac{1}{2} \rho (r) $
for $B>B_*$. 
\begin{figure}[ht]
\centering
\includegraphics*[width=4in]{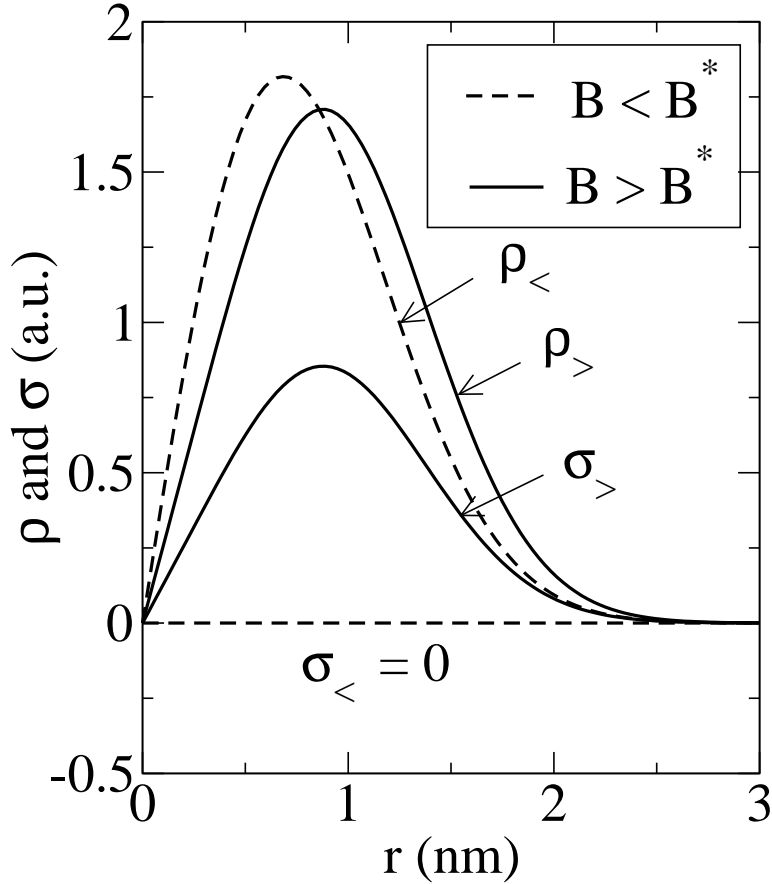}
\vskip.2in
\caption{Charge density as a function of the radius $r$. $B<B_*$ ($B>B_*$) 
should be understood as $B$ slightly smaller (higher) than $B_*$. 
The charge density $\rho$ is only slightly affected when $B$ goes across $B_*$.
The spin density $\sigma$ is zero when $B<B_*$. It has  the profile
of the charge density as $B>B_*$ ($\sigma _> = \met \rho _> $).}
\end{figure}

The dynamics of the dot between the two macroscopic spin states $S=0,1$ is 
induced by the coupling to the contacts. Tunneling to and from the dot is 
only virtual and occupation of the dot, if tuning is appropriate, is still
$N=2$. As shown by a Schrieffer-Wolff  transformation \cite{noi},
 coupling to 
the leads involves both orbital and spin variables. However, in the 
special situation here devised, the exchange of angular momentum is locked 
in with  that of the spin in such a way that  coupling  is overall  of  
AF type. 

This Kondo coupling is peculiar. Of the four levels involved, 
the two crossing levels play 
the role of 
the levels of an effective spin $\met$, $S_{eff}$, which is acted on by 
the deviations $\delta B = B-B_* $, only.  The other two 
levels can be attributed to another spin $\met$, $S_r$   
which is  fully  decoupled.    In the dynamics of $S_{eff}$, the
external magnetic field  $B_*$ disappears completely, provided  conduction 
electrons of both spin orientations  are present at  $\epsilon _F$. If 
 the hybridization with the contacts, $\Delta$, is
large enough and $T < T_K$, the system
flows towards the strongly-coupled  fixed point of a standard 
spin $\met$ Kondo model, unlike what happens at the TS crossing point. 
The spin $S_{eff}$ 
is screened out by the spin density of 
the  delocalized electrons and $S_r$ only survives.  We end up with 
 charge $N=2$ on the dot 
and a dot spin $S_r =1/2$ with levels splitted in the magnetic field.
This is the inverted effect  $w.r.to$  Kondo coupling for $N= \:odd$ and 
 leads to   fractionalization of the spin in the QD.

\section{Summary}
%This is a \inx{summary} of this article.

To conclude, Kondo conduction in a QD  at CB  is the striking  realization 
of a  macroscopic entangled state between the dot and the contacts. 
 Alternatively, it can be seen as an extreme condition by which 
the measuring apparatus is  fully  invasive. Tunneling 
across the dot is no longer perturbative and  separation of the 
dot  total spin  
from the dot charge  sets in. Because of the internal structure 
of the  dot, new properties of the Kondo conduction arise, which are 
not present in  the conductivity of diluted alloys where the Kondo effect 
 was first discovered.  In particular, while the presence of a magnetic 
field, by lifting the degeneracy of the impurity levels is disruptive 
in diluted alloys, strong  Coulomb interaction in dots can give rise to 
drastic orbital changes and to level crossing. In these conditions 
the magnetic field   acts in favour of a Kondo coupling and strong 
conduction anomalies in dots at Coulomb blockade can be measured. 
The Kondo temperature in  these systems is rather low ($T_K \sim 100 mK $ 
and below ). This notwithstanding, it is the very discreteness of the 
levels in the confined dot geometry to support  the flow to strong 
coupling at low temperature, irrespective of the  influence of the 
environment.   

\begin{acknowledgments}

The authors acknowledge useful discussions with  B.Altshuler, S.De Franceschi,
B.Kramer, G.Morandi and  J.Weis.

Work supported by INFM (Pra97-QTMD )
and by EEC with TMR project, contract FMRX-CT98-0180.
\end{acknowledgments}

\end{document}